\documentclass[american,aps,prl,reprint,superscriptaddress,longbibliography]{revtex4-1}
\usepackage[T1]{fontenc}
\usepackage[latin9]{inputenc}
\usepackage{color}
\usepackage{babel}
\usepackage{amsthm}
\usepackage{amsmath}
\usepackage{amssymb}
\usepackage{graphicx}
\usepackage[unicode=true,pdfusetitle,
 bookmarks=true,bookmarksnumbered=false,bookmarksopen=false,
 breaklinks=false,pdfborder={0 0 0},backref=false,colorlinks=false]
 {hyperref}
\hypersetup{
 colorlinks,linkcolor=myurlcolor,citecolor=myurlcolor,urlcolor=myurlcolor}

\makeatletter
\@ifundefined{textcolor}{}
{%
 \definecolor{BLACK}{gray}{0}
 \definecolor{WHITE}{gray}{1}
 \definecolor{RED}{rgb}{1,0,0}
 \definecolor{GREEN}{rgb}{0,1,0}
 \definecolor{BLUE}{rgb}{0,0,1}
 \definecolor{CYAN}{cmyk}{1,0,0,0}
 \definecolor{MAGENTA}{cmyk}{0,1,0,0}
 \definecolor{YELLOW}{cmyk}{0,0,1,0}
 }
\theoremstyle{plain}

\theoremstyle{plain}

\ifx\proof\undefined
\newenvironment{proof}[1][\protect\proofname]{\par
\normalfont\topsep6\p@\@plus6\p@\relax
\trivlist
\itemindent\parindent
\item[\hskip\labelsep
\scshape
#1]\ignorespaces
}{%
\endtrivlist\@endpefalse
}
\providecommand{\proofname}{Proof}
\fi
\theoremstyle{plain}

\providecommand{\lemmaname}{Lemma}
\providecommand{\definitionname}{Definition}
\providecommand{\propositionname}{Proposition}

\usepackage{babel}
\usepackage{txfonts}%\usepackage{braket}
\usepackage{colortbl}\definecolor{myurlcolor}{rgb}{0,0,0.7}

\newcommand{\norm}[1]{\left\| #1 \right\|}
\newcommand{\tr}{{\operatorname{Tr\,}}}
\newcommand{\id}{{\mathbb{I}}}
\def\ket#1{| #1 \rangle}
\def\bra#1{\langle  #1 |}

\newcommand{\emphasize}[1]{{\color{black}#1}}

\newcommand\Sys{\mathrm{S}}
\newcommand\B{\mathrm{B}}
\newcommand\SB{\mathrm{SB}}
\newcommand\AB{\mathrm{AB}}
\newcommand\A{\mathrm{A}}
\newcommand\AC{\mathrm{AC}}

\providecommand{\lemmaname}{Lemma}
\providecommand{\propositionname}{Proposition}
\providecommand{\theoremname}{Theorem}

\begin{document}

\title{Generalized Laws of Thermodynamics in the Presence of Correlations}

\author{Manabendra N. Bera}
\email{Manabendra.Bera@icfo.eu}
% \selectlanguage{american}%

\affiliation{ICFO -- Institut de Ci\`encies Fot\`oniques, The Barcelona Institute of Science and Technology, ES-08860 Castelldefels, Spain}

\author{Arnau Riera}
\affiliation{ICFO -- Institut de Ci\`encies Fot\`oniques, The Barcelona Institute of Science and Technology, ES-08860 Castelldefels, Spain}
\affiliation{Max-Planck-Institut f\"ur Quantenoptik, D-85748 Garching, Germany}

\author{Maciej Lewenstein} 

\affiliation{ICFO -- Institut de Ci\`encies Fot\`oniques, The Barcelona Institute of Science and Technology, ES-08860 Castelldefels, Spain}

\affiliation{ICREA -- Instituci\'o Catalana de Recerca i Estudis Avan\c{c}ats, Pg.~Lluis Companys 23, ES-08010 Barcelona, Spain}

\author{Andreas Winter}

\affiliation{ICREA -- Instituci\'o Catalana de Recerca i Estudis Avan\c{c}ats, Pg.~Lluis Companys 23, ES-08010 Barcelona, Spain}

\affiliation{Departament de F\'{i}sica: Grup d'Informaci\'o Qu\`antica, Universitat Aut\`onoma de Barcelona, ES-08193 Bellaterra (Barcelona), Spain}

% \date{11 December 2016}

\maketitle

\noindent {\bf Abstract}
\vspace{.3cm}

The laws of thermodynamics, despite their wide range of applicability, are known to break
down when systems are correlated with their environments.
Here, we generalize thermodynamics to physical scenarios which allow presence of correlations, 
including those where strong correlations are present. 
We exploit the connection between information and physics, 
and introduce a consistent redefinition of heat dissipation by systematically accounting 
for the information flow from system to bath 
in terms of the conditional entropy.
As a consequence, the formula for the Helmholtz free energy is accordingly modified.
Such a remedy not only fixes the apparent
violations of Landauer's erasure principle and the second law due to anomalous 
heat flows, 
but also leads to a generally valid reformulation of the laws of thermodynamics.
In this information-theoretic approach, correlations between system and environment store
work potential. Thus, in this view, the apparent anomalous heat flows are the refrigeration processes 
driven by such potentials.

\vspace{.7cm}
\noindent  {\bf Introduction}
\vspace{.3cm}

Thermodynamics is one of the most successful physical theories ever formulated. 
Though it was initially developed to deal with steam engines and, in particular, 
the problem of conversion of heat into mechanical work,
it has survived even after the scientific revolutions of relativity and quantum mechanics. 
Inspired by resource theories, 
%\cite{Horodecki13a, Bennett96, Horodecki09, Bartlett07, Horodecki03, Baumgratz14, Winter16}, 
recently developed in quantum information,
% \cite{Shannon48, Nielsen00, Cover05}, 
a renewed effort has been made to understand the foundations of thermodynamics 
in the quantum domain
% concerning thermodynamic work extraction 
\cite{Brandao13, Horodecki13, Aberg13, Faist15, Skrzypczyk14, Brandao15, Cwiklinski15, Lostaglio15, Lostaglio15a, Dahlsten11, Egloff15}, including its connections to statistical mechanics \cite{Gogolin16, Short11, Popescu06} and information theory \cite{Reeb14, Maruyama09, Bennett82, Parrondo15, Leff90, Leff02, Landauer61, Szilard29, Goold16, Rio11, Marti15}. 
However, all these approaches assume that the system is initially uncorrelated from the bath. 
In fact, in the presence of correlations, 
the laws of thermodynamics can be violated. 
In particular, when there are inter-system correlations,
phenomena such as anomalous heat flows from cold to hot baths \cite{Jennings10}, 
and memory erasure accompanied by work extraction instead of heat dissipation \cite{Rio11} become possible. 
These two examples indicate a violation of the second law in its Clausius formulation,
and the Landauer's principle of information erasure \cite{Reeb14} respectively.
Due to the interrelation between the different laws of thermodynamics, 
the zeroth law and the first law can also be violated (see Supplementary Note 4 for simple and explicit examples of these violations).

The theory of thermodynamics can be summarized in its three main laws.
The \emphasize{zeroth law} introduces the notion of thermal \emphasize{equilibrium} 
as an equivalence relation of states,
where \emphasize{temperature} is the parameter that labels the different equivalence classes.
In particular, the transitive property of the equivalence relation implies 
that if a body A is in equilibrium with a body B, and
B is with a third body C, then A and C are also in equilibrium.
The \emphasize{first law} assures energy conservation. 
It states that in a thermodynamic process not all of energy changes are of the same
nature and distinguishes between \emphasize{work}, the type of energy that allows 
for ``useful'' operations as raising a weight, and its complement \emphasize{heat}, any 
energy change which is not work.
Finally, the \emphasize{second law} establishes an arrow of time. 
It has several formulations and perhaps the most common one is the \emphasize{ Clausius} statement, 
which reads:
\emphasize{ No process is possible whose sole result is the 
transfer of heat from a cooler to a hotter body.}
Such a restriction not only introduces the fundamental limit on how and to what extent various forms of energy can be converted to accessible mechanical work, but also implies the existence of an additional state function, the \emphasize{entropy}, which has to increase. 
There is also the \emphasize{third law} of thermodynamics; we shall, however,
leave it out of the discussion, 
as it is beyond immediate context of the physical scenarios considered here.

Although the laws of thermodynamics were developed phenomenologically, they have
profound implications in information theory. The paradigmatic example is the Landauer erasure principle,
which states: 
\emphasize{``Any logically irreversible manipulation of information, such as the erasure of a bit or the merging of two computation paths, must be accompanied by a corresponding entropy increase in non-information-bearing degrees of freedom of the information-processing apparatus or its environment''} \cite{Bennett82}. Therefore, an erasing operation is bound to be associated with a heat flow to the environment.

An important feature in the microscopic regime is that the quantum particles can exhibit non-trivial correlations, such as entanglement \cite{Horodecki09}
and other quantum correlations \cite{Modi12}. 
Thermodynamics in the presence of correlations has been considered only 
in limited physical situations. It is assumed, in nearly all cases of 
thermodynamical processes, that system and bath are initially uncorrelated,
although correlations may appear in the course of the process. 
In fact, it has been noted that in the presence of such correlations,
Landauer's erasure principle could be violated \cite{Reeb14}. 
Even more strikingly, with strong quantum correlation between two thermal 
baths of different temperatures, heat could flow from the colder bath to 
the hotter one \cite{Jevtic12, Jennings10, Partovi08}.

The impact of inter-system correlations resulting from a strong system-bath coupling and its role in thermodynamics has been studied for some specific solvable models \cite{Allahverdyan00, Ford06, Hilt11}, and for general classical systems \cite{Seifert16, Jarzynski17}. 
It has been noted that presence of correlations requires certain adjustments of work and heat to fulfil the second law and the Landauer principle. Also, from an information theoretic perspective, both extractable work from correlations and work cost to create correlations have been studied \cite{Marti15, Huber15, Bruschi15, Friis16}. However, in all these works, there is no explanation of how to deal with general correlated scenarios irrespective of 
 where the correlations come from and in systems away from
 thermal equilibrium.

Here we show that the violations of the laws of thermodynamics (see Supplementary Note 4) indicate 
that correlations between two systems, irrespective of the corresponding
marginals being thermal states or not, manifest out-of-equilibrium phenomena.
In order to re-establish the laws of thermodynamics, one not only has to look 
at the local marginal systems, but also the correlations between them. 
In particular, we start by redefining the notions of heat and work, then establish a generalized Landauer's principle 
and introduce the generalized Helmholtz free-energy.
The resulting laws are  \emphasize{general} in the sense that they rely on the least set of assumptions to formulate thermodynamics: a system, 
a considerably large thermal bath at well defined temperature, and separable initial and final Hamiltonians. 
The first two assumptions are obvious. 
The third assumption is basically required for system's and bath's energies to be well defined (see Supplementary Note 2 for details).

\vspace{.7cm}
\noindent {\bf Results}
\vspace{.3cm}

\noindent {\bf Definition of heat}

\noindent
To reformulate thermodynamics, we start with redefining heat by properly accounting for the information 
flow and thereby restoring Landauer's erasure principle. 
In general, \emphasize{heat} is defined as the flow of energy from the environment,
normally considered as a thermal \emphasize{ bath} at certain temperature, 
to a system, in some way different from \emphasize{ work}.
Work, on the other hand, is quantified as \emphasize{the flow of energy, say to a 
bath or to an external agent, that could be extractable (or accessible)}. 
% Operationally, it is energy that can be used to raise a weight or similar
% battery system.
% Therefore, the difference between change in internal energy and extractable 
% work quantifies the heat flow to the bath. 
Consider a thermal bath with Hamiltonian $H_\B$ and at temperature $T$ represented by 
the Gibbs state $\rho_\B=\tau_\B=\frac{1}{Z_\B}\exp(\frac{-H_\B}{kT})$, where $k$ is the Boltzmann constant,
and $Z_\B=\mbox{Tr}\left[\exp(\frac{-H_\B}{k T}) \right]$ is the partition function. 
The degrees of freedom in B are considered to be a part of a large thermal super-bath, at temperature $T$.
Then, for a process that transforms the thermal bath $\rho_\B\rightarrow \rho_\B^\prime$ with the fixed Hamiltonian $H_\B$, the heat transfer to the bath is quantified (see Supplementary Note 1) as 
\begin{equation}
\label{eq:defHeat}
 \Delta Q = -kT\, \Delta \mathcal{S}_\B,
\end{equation}
where $\Delta \mathcal{S}_\B=\mathcal{S}(\rho_\B^\prime)-\mathcal{S}(\rho_\B)$ is 
the change in bath's von Neumann entropy, 
$\mathcal{S}(\rho_\B)=-\mbox{Tr}\left[\rho_\B \log_2 \rho_\B \right] $.
Note that $\rho_\B^\prime$ is not in general thermal.
In fact, the work stored in the bath is $\Delta F_\B$, where $F(\rho_\B)=E(\rho_\B)-kT \mathcal{S}(\rho_\B)$ 
is the Helmholtz free energy, with $E(\rho_\B)=\tr (H_\B \rho_\B)$. 
Heat expressed in Eq.~\eqref{eq:defHeat} is the correct quantification of heat (for further discussion see Supplementary Note 1), 
which can be justified in two ways.
First, it has a clear information-theoretic interpretation, 
which accounts for the information flow to the bath. 
Second, it is the flow of energy to the bath other than work and, 
with the condition of entropy preservation, any other form of 
energy flow to the bath will be stored as extractable work, 
and thus will not converted into heat.
The process-dependent character of heat as defined here can be seen from the fact
that it cannot be written as a difference of state functions of the system. 
In the Supplementary Note 1, this issue is discussed and the sources of irreversibility, 
i.e. the reasons for not saturating the Clausius inequality, are re-examined.

The transformations considered in our framework are
\emphasize{entropy preserving} operations. More explicitly, given a system-bath
setting initially in a state $\rho_\SB$, in which the reduced state of the system $\rho_\Sys$ is arbitrary
while $\rho_\B$ is thermal,
we consider transformations 
$\rho_\SB'=\Lambda(\rho_\SB)$ such that the von Neumann entropy is unchanged
i.~e.~$S(\rho_\SB')=S(\rho_\SB)$.
The Hamiltonians of the system and the bath are the same before 
and after the transformation $\Lambda(\cdot)$.
Note that we do not demand energy conservation, rather assuming that a suitable battery
takes care of that. In fact, the work cost of such an operation $\Lambda(\cdot)$ is
quantified by the global internal energy change
$\Delta W=\Delta E_\Sys + \Delta E_\B$.
Another comment to make is that we implicitly assume a bath of unbounded
size; namely, it consists of the part $\rho_\B$ of which we explicitly
track the correlations with S, but also of arbitrarily many independent
degrees of freedom. Also, we are implicitly considering always the
asymptotic scenario of $n\rightarrow\infty$ copies of the state in question
(``thermodynamic limit'').
These operations are general and include any process and situation in standard thermodynamics involving a single bath. It is the result of abstracting the essential elements of thermodynamic processes: 
existence of a thermal bath and global entropy preservation operations.

In extending thermodynamics in correlated scenarios and linking thermodynamics with information, 
we consider the \emphasize{ quantum conditional entropy}
as the natural quantity to represent information content in the system as well as 
in the correlations.
For a joint system-bath state $\rho_\SB$, the information content in the system 
S, given all the information available in the bath B at temperature $T$, 
is quantified by the conditional entropy 
$\mathcal{S}(\Sys |\B)=\mathcal{S}(\rho_\SB)-\mathcal{S}(\rho_\B)$. 
It vanishes when the joint system-environment state is perfectly classically correlated
and can even become negative in the presence of entanglement. 

\

\noindent {\bf Generalized second law of information}

\noindent
With quantum conditional entropy, 
the generalized second law of information can be stated as follows.
For an entropy preserving operation $\rho_\SB^\prime=\Lambda^\SB (\rho_\SB)$, 
with the reduced states before (after) the evolution denoted
$\rho_\Sys$ ($\rho_\Sys^\prime$) and $\rho_\B$ ($\rho_\B^\prime$), respectively, we have
\begin{align}
  \Delta \mathcal{S}_\B=-\Delta \mathcal{S}(\Sys |\B),
\end{align}
where $\Delta \mathcal{S}_\B=\mathcal{S}(\rho_\B^\prime)-\mathcal{S}(\rho_\B)$ is 
the change in (von Neumann) entropy of the bath, 
and $\Delta \mathcal{S}(\Sys |\B)=\mathcal{S}(S^\prime|B^\prime)-\mathcal{S}(\Sys |\B)$ 
is the change in conditional entropy of the system. 
Note that in the presence of initial correlations, 
the informational second law could be violated if one considers only system entropy 
(see Supplementary Note 3).

Let us point out 
that the conditional entropy of the system for a given bath is also used in \cite{Rio11} in the context
of erasing. 
There, it is shown that the conditional entropy quantifies the amount of work necessary to erase quantum information. The formalism in \cite{Rio11} considers energy preserving but non-entropy preserving operations and that perfectly enables to  quantify work. In contrast, in our formalism, as we attempt to quantify heat in connection with information flow, it is absolutely necessary to guarantee information conservation, thereby restrict ourselves to entropy preserving operations. This leads us to quantify heat in terms of conditional entropy. Both approaches are different and complement each other. In one, the conditional entropy quantifies work and on the other, it quantifies heat.

\

\noindent {\bf Generalized Landauer's principle}

\noindent
The  Landauer principle is required to be expressed in 
terms of conditional entropy of the system, rather than its local entropy.
% , and is stated in Theorem \ref{thm:1}.
Therefore, the dissipated heat associated to information erasure of a system S connected to a bath 
B at temperature $T$ by an entropy preserving operation $\rho_\SB^\prime=\Lambda^\SB (\rho_\SB)$, is equal to
\begin{equation}
\label{eq:uL}
 \Delta Q = kT\, \Delta \mathcal{S}(\Sys |\B)\, .
\end{equation}
Note that, in complete information erasure, the final conditional
entropy vanishes, then $\Delta Q =  - kT\,  \mathcal{S}(\Sys |\B)$. 

\

\noindent {\bf Generalized Helmholtz free-energy}

\noindent
We address extraction of work from a system S possibly correlated to a bath B at 
temperature $T$. Without loss of generality, we assume that the system Hamiltonian 
$H_\Sys$ is unchanged in the process. Note that the extractable work has two contributions:
one comes from system-bath correlations (cf.~\cite{Marti15}) and the other from the local system alone,
irrespective of its correlations with the bath. Here we consider these two contribution separately. 

\begin{figure}[h] 
\includegraphics[scale=0.3]{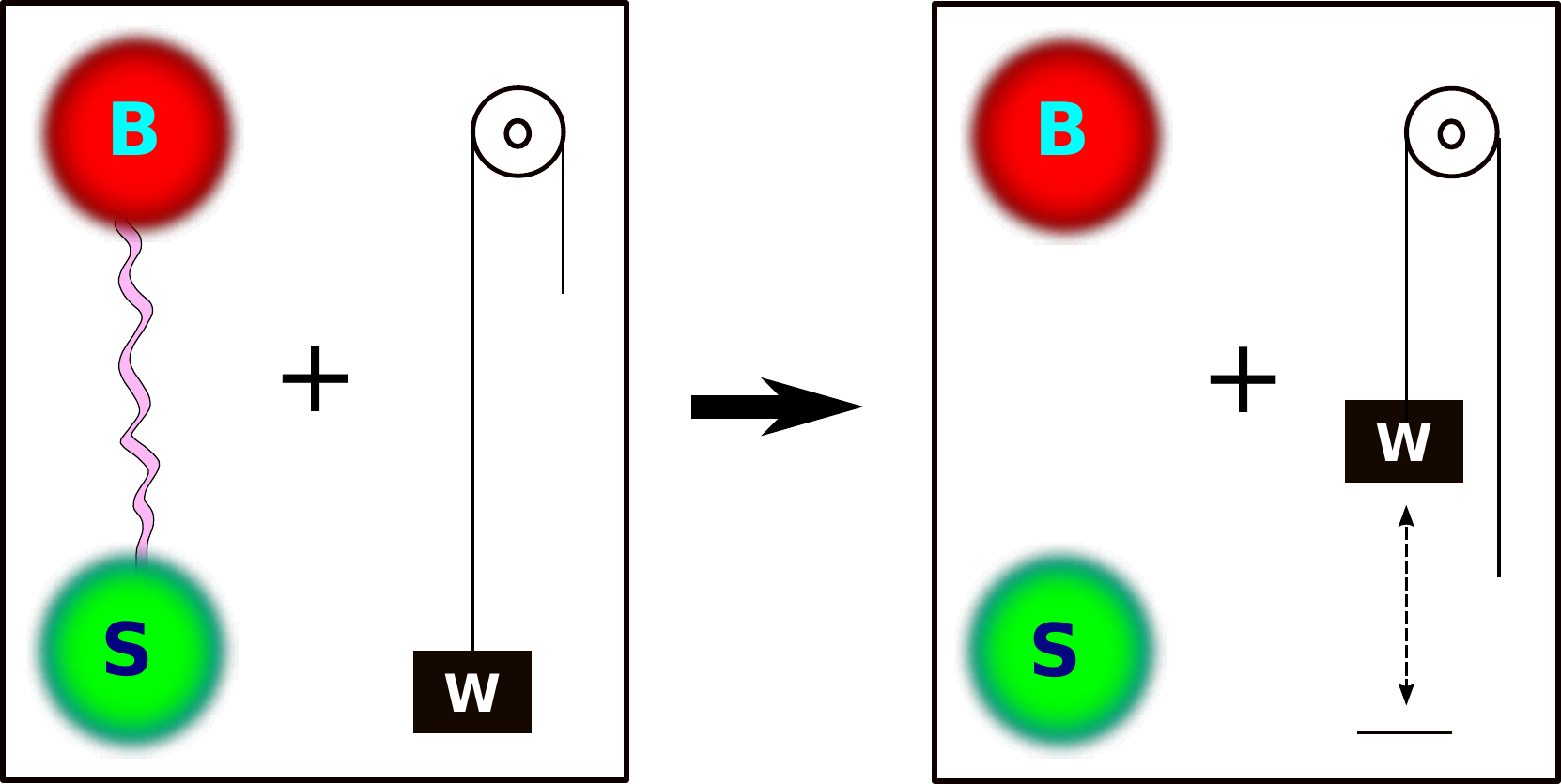}
\caption{\textbf{Correlations as a work potential.} Correlations can be understood as a work potential, as quantitatively expressed in Eq.~\eqref{eq:CtoW}.}
\label{fig:CtoW}
\end{figure}

By extracting work from the correlation, we mean any process that returns the system and the bath in the original reduced states, $\rho_\Sys$ and $\rho_\B = \tau_\B$. The maximum extractable work solely from the correlation, using entropy preserving operations, is given by
\begin{align}\label{eq:CtoW}
 W_C= kT\, \mathcal{I}(\Sys :\B),
\end{align}
where $\mathcal{I}(\Sys :\B)= \mathcal{S}_\Sys + \mathcal{S}_{B} - \mathcal{S}_\SB$ is the mutual information. 
This is illustrated in Fig.~\ref{fig:CtoW}.
The proof is given by the protocol described in {\bf Box 1}.

\begin{tabular}{| p{8cm}| }
  \hline
  \vspace{.1cm}			
  \hspace{.1cm} {\bf Box 1 | Work extraction from correlations. } \vspace{.2cm}\\
  \hline
{\small
\noindent{\bf Addition of an ancillary system}: 
We attach to $\rho_\SB$ an ancillary system A with trivial 
Hamiltonian $H_\A=0$, consisting of $\mathcal{I}(\Sys :\B)$ qubits in the
maximally mixed state $\tau_\A = \left(\frac{\mathbb{I}_2}{2}\right)^{\otimes \mathcal{I}(\Sys :\B)}$
(which is thermal!).

\noindent{\bf Removing the correlations between S and B}: 
By using a global entropy preserving operation, we make a transformation $\tau_{A}\otimes\rho_\SB \rightarrow \tau_\A^\prime\otimes\rho_\Sys\otimes\rho_\B$, where 
\begin{equation}
\mathcal{S}(\A\Sys|\B)_{\tau_\A \otimes \rho_\SB} 
   = \mathcal{S}(\A^\prime \Sys|\B)_{\tau_\A^\prime \otimes \rho_\Sys \otimes \rho_\B},
\end{equation}
and thereby turning the additional state into a pure state $\tau_\A^\prime=\ket{\phi}\bra{\phi}$ of A,
while leaving the marginal system and bath states unchanged. 
Clearly, the extractable work stored in the correlation is now transferred 
to the new additional system state $\tau_\A^\prime$.

\noindent{\bf Work extraction}: Work is extracted from $\tau_\A^\prime$ at temperature $T$, 
equal to $W_C= \mathcal{I}(\Sys :\B)_{\rho_\SB} \ kT$.}
\\
  \hline  
\end{tabular}
%
%\begin{itemize}
%\item Step 1. We attach to $\rho_\SB$ an ancillary system A with trivial 
%Hamiltonian $H_\A=0$, consisting of $\mathcal{I}(\Sys :\B)$ qubits in the
%maximally mixed state $\tau_\A = \left(\frac{\mathbb{I}_2}{2}\right)^{\otimes \mathcal{I}(\Sys :\B)}$
%(which is thermal!).
%\item Step 2. By using a global entropy preserving operation, we make a transformation $\tau_{A}\otimes\rho_\SB \rightarrow \tau_\A^\prime\otimes\rho_\Sys\otimes\rho_\B$, where 
%\begin{align}
%\mathcal{S}(\A\Sys|\B)_{\tau_\A \otimes \rho_\SB} 
%   = \mathcal{S}(\A^\prime \Sys|\B)_{\tau_\A^\prime \otimes \rho_\Sys \otimes \rho_\B},
%\end{align}
%and thereby turning the additional state into a pure state $\tau_\A^\prime=\ket{\phi}\bra{\phi}$ of A,
%while leaving the marginal system and bath states unchanged. 
%Clearly, the extractable work stored in the correlation is now transferred 
%to the new additional system state $\tau_\A^\prime$.  
%\item Step 3. Work is extracted from $\tau_\A^\prime$ at temperature $T$, 
%equal to $W_C= \mathcal{I}(\Sys :\B)_{\rho_\SB} \ kT$.
%\end{itemize}

\vspace{.5cm}
Disregarding the correlations with a bath at temperature $T$, the maximum extractable work from a state $\rho_\Sys$ is given by 
$\Delta W_L=F(\rho_\Sys)-F(\tau_\Sys)$, where $\tau_\Sys=\frac{1}{Z_\Sys}\exp[- \frac{H_\Sys}{kT}]$ 
is the corresponding thermal state of the system in equilibrium with the bath. 
Now, in addition to this ``local work'', we have the work due to correlations,
and so the total extractable work $\Delta W_\Sys=\Delta W_L + kT\, \mathcal{I}(\Sys :\B)_{\rho_\SB}$.
Note that, for the system alone, the Helmholtz free energy 
$F(\rho_\Sys)=E_\Sys - kT\, \mathcal{S}_\Sys$. However, in the presence of correlations, 
it is modified to \emphasize{generalized Helmholtz free energy}, by adding $kT\, \mathcal{I}(\Sys :\B)_{\rho_\SB}$ to $F(\rho_\Sys)$, as
\begin{align}
 \label{eq:modified-Helmotz}
 \mathcal{F}(\rho_\SB)= E_\Sys - kT\, \mathcal{S}(\Sys |\B).
\end{align}
Unlike the traditional free energy, the generalized free energy
is not only a state function of the system S, 
but also of those degrees of freedom of the bath correlated with it.
This is an unavoidable feature of the generalised formalism.
Therefore, for a system-bath state $\rho_\SB$,
maximum extractable work from the system can be given as 
$\Delta W_\Sys=\mathcal{F}(\rho_\SB) - \mathcal{F}(\tau_\Sys\otimes\tau_\B)$, where 
$\mathcal{F}(\tau_\Sys\otimes\tau_\B)= F(\tau_\Sys)$. Then, for a transformation, 
for which initial and final states are $\rho_\SB$ and $\sigma_\SB$, respectively, the maximum extractable work from the system, 
is $\Delta W_\Sys= -\Delta \mathcal{F}=\mathcal{F}(\rho_\SB)-\mathcal{F}(\sigma_\SB)$.
We observe that all this is of course consistent with what we know from 
situations with an uncorrelated bath. Indeed, we can simply make the conceptual step of
calling SB ``the system'', allowing for arbitrary correlations
between S and B, with a suitable infinite bath B' that is uncorrelated
from SB. Then, the free energy as we know it is $F(\rho_\SB)=E_\Sys - kT\, \mathcal{S}(\Sys |\B) + E_\B - kT\, \mathcal{S}(\tau_\B)$,
% \begin{align}\begin{split}
%   F(\rho_\SB) &= E_\Sys + E_\B - kT\, \mathcal{S}(\rho_\SB) \\
%                &= E_\Sys - kT\, \mathcal{S}(\Sys |\B) + E_\B - kT\, \mathcal{S}(\tau_\B), 
% \end{split}\end{align}
where the first term is the modified free energy in Eq. \eqref{eq:modified-Helmotz},
and the second term is the free energy of the bath in its thermal state.
As the latter cannot become smaller in any entropy-preserving operation,
the maximum extractable work is $-\Delta\mathcal{F}$.

\

\noindent {\bf Generalized laws of thermodynamics}

\noindent
Now, equipped with the proper definition of heat (as in Eq.~\eqref{eq:uL}) and work (based on generalized free energy in Eq.~\eqref{eq:modified-Helmotz}) in the presence of correlations, we put forward the \emphasize{generalized laws of thermodynamics}.

We start with \emphasize{generalized first law}, which states: \emphasize{given an entropy preserving operation $\rho_\SB \rightarrow \rho_\SB^\prime $, 
%  where the local system state follow the transformation $\rho_\Sys \rightarrow\rho_\Sys^\prime$, 
 the distribution of the change in the system's internal energy into work and heat satisfies
 \begin{align}
 \label{eq:1stLaw}
  \Delta E_\Sys=-(\Delta W_\Sys + \Delta F_\B) +(\Delta Q + \Delta F_\B ),
 \end{align}
 where the heat dissipated to the bath is given by $\Delta Q= - kT\, \Delta \mathcal{S} (\Sys |\B)$, the maximum extractable work from the system is $\Delta W_\Sys =- \left( \Delta E_\Sys - kT\, \Delta \mathcal{S} (\Sys |\B) \right)$, and the work performed on the bath is $\Delta F_\B = \Delta E_\B - kT\, \Delta \mathcal{S}_\B\geqslant 0$.}

The quantity $\Delta W_\Sys=-\left(\Delta E_\Sys - kT\, \Delta \mathcal{S} (\Sys |\B) \right)$ was shown to be the maximum extractable work, as it is equal to $-\Delta \mathcal{F}_\Sys$. The maximum work $\Delta W_\Sys$ is extracted by thermodynamically reversible processes. 
Irreversible processes require that some work is performed on the bath $\Delta F_\B > 0$
followed by an equilibration process, which happens due to spontaneous relaxation of the bath. Such amount of work is transformed into heat and hence cannot be accessed any more. 
Note that such an equilibration process is not entropy preserving \cite{Gogolin16} which is not allowed in our setup.
The entropy production of such relaxation is precisely $\Delta F_\B /T$, and in that case heat flow from the bath is exactly equal to the decrease of its internal energy.

\begin{figure}[h]
\includegraphics[scale=.5]{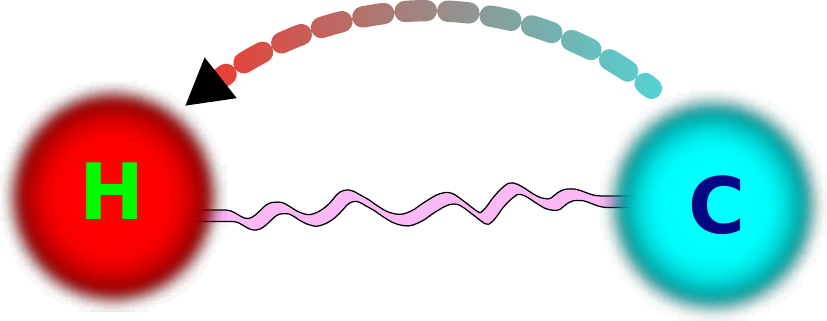}
\caption{\textbf{Anomalous heat flows.} In the presence of correlations, spontaneous heat flows from cold to hot baths are possible \cite{Jennings10}. This is an apparent violation of second law, if one ignores the work potential stored in correlation. Otherwise, it is a refrigeration process.}
\label{fig:anomalous-heat}
\end{figure}
In this new approach, the second law is also modified. The  
%  \item \emphasize{Kelvin-Planck statement of the generalized second law:} No process is possible whose sole result is the absorption of heat from a reservoir and the conversion of this heat into work, where heat and work are defined as in Eq.~\eqref{eq:1st-a} and \eqref{eq:1st-b} respectively.
\emphasize{Clausius statement of the generalized second law} states that \emphasize{no process is possible whose sole result is the transfer of heat from a cooler to a hotter body, where the work potential stored in the correlations, as defined in Eq.~\eqref{eq:CtoW}, does not decrease}.
To prove it, consider a state transformation
$\rho_\AB^\prime=\Lambda^\AB (\rho_\AB)$ where $\Lambda^\AB$ is an entropy preserving and
and energy non-increasing operation.
As the thermal state minimizes the free energy,  the final reduced states $\rho_\Sys^\prime$ and $\rho_\B^\prime$  have increased their free energy, i.e., $\Delta E_\A -k T_\A \Delta \mathcal{S}_\A \geqslant 0$ and $\Delta E_\B -k T_\B \Delta \mathcal{S}_\B \geqslant 0$, 
where $T_{\A / \B}$, $\Delta E_{\A/ \B}$ and $\Delta \mathcal{S}_{\A/ \B}$ are the initial temperatures, changes in internal energy and entropy of the baths respectively.
% \begin{align}
% \Delta E_\A -k T_\A \Delta \mathcal{S}_\A &\geqslant 0 \label{eq:free-energy-increaseA}\\
% \Delta E_\B -k T_\B \Delta \mathcal{S}_\B &\geqslant 0\, ,\label{eq:free-energy-increaseB}
% \end{align}
% where $T_{A/B}$ is the initial temperature of the baths, and $\Delta E_{A/B}$ and $\Delta S_{A/B}$ are the change in internal energy and entropy respectively.
By adding the former inequalities and considering energy non-increasing, we get $T_\A \Delta \mathcal{S}_\A + T_\B \Delta \mathcal{S}_\B \leqslant 0$.
% Eqs.~\eqref{eq:free-energy-increaseA} and \eqref{eq:free-energy-increaseB}, and considering energy conservation, we get
% \begin{equation}\label{eq:Clausius-step}
% T_\A \Delta \mathcal{S}_\A + T_\B \Delta \mathcal{S}_\B \le 0\, .
% \end{equation}
Due to the conservation of total entropy,
the change in mutual information is simply $\Delta I(\A :\B)=\Delta \mathcal{S}_\A+ \Delta \mathcal{S}_\B$,
with $I(\A :\B)=\mathcal{S}_\A + \mathcal{S}_\B - \mathcal{S}_\AB$. 
This allows us to conclude
\begin{equation}\label{eq:refrigerator}
-\Delta Q_\A (T_\B - T_\A) \geqslant k T_\A T_\B \Delta I(\A:\B),
\end{equation}
% where we have used the work potential given in Eq.~\eqref{eq:CtoW}.
% Equation \eqref{eq:refrigerator} 
which implies \emphasize{Clausius statement of the generalized second law}. 

Note that if the initial state $\rho_\AB$ is correlated, then the change in mutual information could be negative, $\Delta I(\A :\B) \leqslant 0$, and $-\Delta Q_\A (T_\B - T_\A) \leqslant 0$.
Note that for $T_\A\leqslant T_\B$ and $\Delta I(\A :\B)\leqslant 0$,
there could be a heat flow from the cold to the hot bath $\Delta Q_\A \geqslant 0$, 
i.e., an apparent anomalous heat flow. 
From our new perspective, we interpret the anomalous heat flow as a refrigeration driven by
the work potential stored in correlations. 
In this case, it
is interesting to determine its coefficient of performance $\eta_\textrm{cop}$, that from Eq.~\eqref{eq:refrigerator} leads, with the work performed on the hot bath
$\Delta W_C(T_\B)=-k T_\B \Delta I(\A :\B)$, to
\begin{equation}\label{eq:refrigeration-cop}
\eta_\textrm{cop}\coloneqq\frac{\Delta Q_\A}{\Delta W_C(T_\B)} \leqslant \frac{T_\A}{T_\B-T_\A}
\end{equation}
which is nothing else than the \emphasize{Carnot coefficient of performance} (see Fig.~\ref{fig:anomalous-heat}).
Note that we have taken the work value of the correlations $W_C$ with respect to the hot bath
$T_\B$. This is due to the fact that for this refrigeration process the hot bath is the one acting as a reservoir.

Equation \eqref{eq:refrigeration-cop} is a nice reconciliation with traditional thermodynamics. 
The Carnot coefficient of performance is a consequence of the fact that reversible processes are optimal, otherwise
the perpetual mobile could be build by concatenating a "better" process and a reversed reversible one. Hence, it is natural that the refrigeration process driven by the work stored in the correlations preserves Carnot statement of second law.

\begin{figure}[h]
\includegraphics[scale=.2]{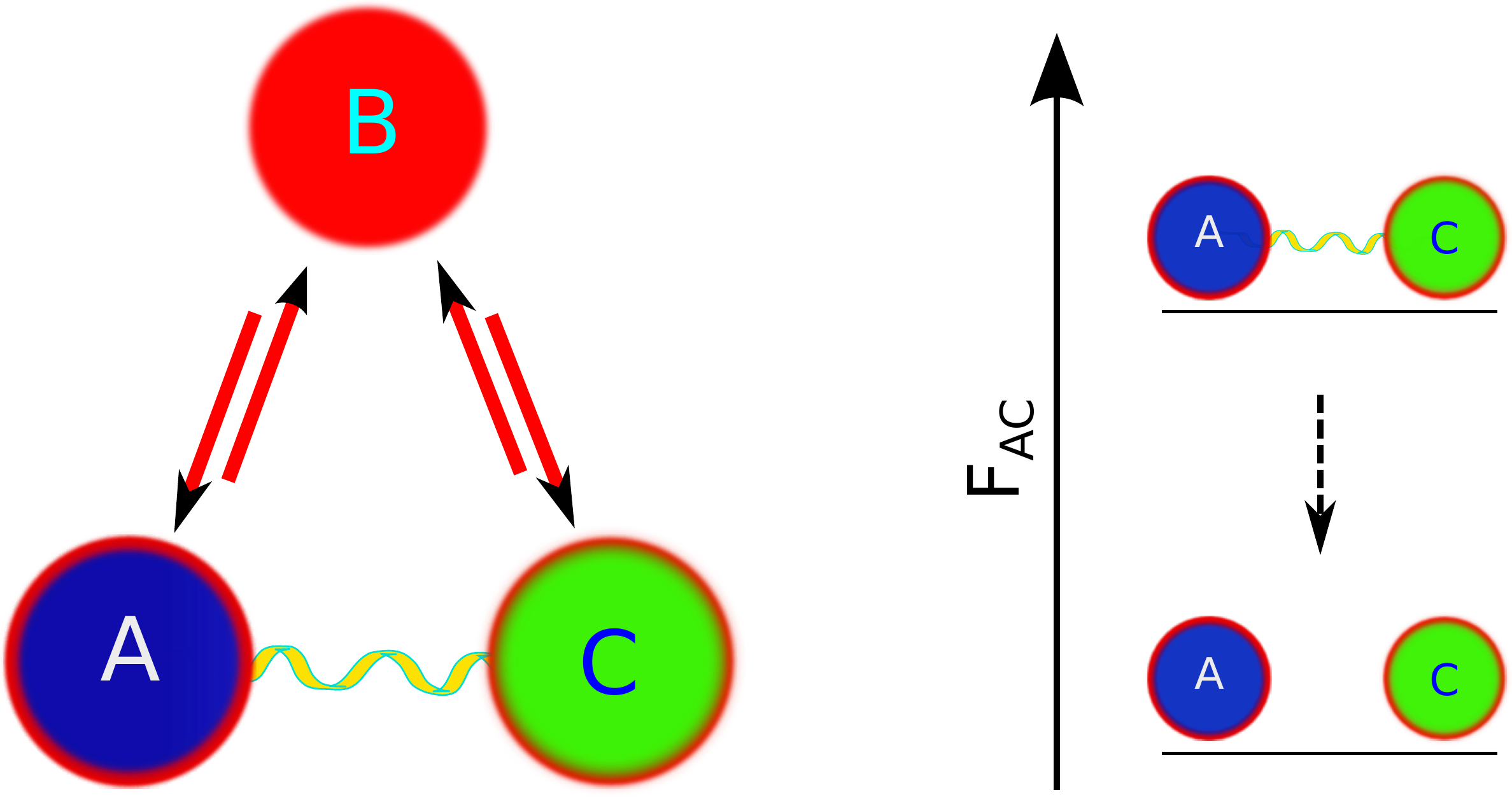}
\caption{\textbf{Violation of the zeroth law.} 
In the presence of correlations, the notion of
equilibrium is not an equivalence relation. 
Consider 3-party state $\rho_\B\otimes\rho_\AC$ with all marginals thermal states.
The thermal equilibria 
A $\leftrightharpoons$ B and B $\leftrightharpoons$ C implies that A, B and C share the same temperature. But, in the presence of correlations between A and C, that does not assure 
the equilibrium A $\leftrightharpoons$ C.
Therefore, the transitive property of the equivalence relation is violated. This is justified, on the right, as $F(\rho_\AC) > F(\rho_\A\otimes\rho_C)$. Thus, the \emphasize{generalized} zeroth law has to overcome these limitations.}
\label{fig:zeroth-law}
\end{figure}

Now, we reconstruct the zeroth law which can be violated in the presence of correlations as shown in Fig.~\ref{fig:zeroth-law}. To do this, we redefine the notion of equilibrium 
beyond an equivalence relation when
correlations between systems are present. Thus, the \emphasize{Universal zeroth law} states that, \emphasize{a collection $\{\rho_X \}_X$ of states is said to be in mutual thermal 
equilibrium with each other if and
only if no work can be extracted from any of their combinations 
under entropy preserving operations.}
This is the case if and only if all the parties $X$ are uncorrelated and each of them is in a thermal state with the same temperature.

\vspace{.7cm}
\noindent {\bf Discussion}
\vspace{.3cm}

Landauer exorcised Maxwell's demon and saved the second law of thermodynamics by taking into account the work potential of information.
In this work, we extend this idea to include also the information about the system 
that is stored in its correlations with the environment. 
With this approach, we easily resolve the apparent violations of thermodynamics 
in correlated scenarios, and generalize it by reformulating its zero-th, first, and second laws.

An important remark is that, 
our generalized thermodynamics is formulated in the asymptotic limit
of many copies. A relevant question is how the laws of thermodynamics are
expressed for a single system. 
In our forthcoming paper, 
% \cite{OneShotUT},
 we will address these questions by discussing consistent notions
 of one-shot heat, one-shot Landauer erasure, and of one-shot
 work extraction from correlations.

\vspace{.7cm}
\noindent \textbf{Acknowledgement}
\vspace{.3cm}

We thank 
R. Alicki, R. B. Harvey, K. Gawedzki, P. Grangier, J. Kimble, V. Pellegrini, R. Quidant, D. Reeb, F. Schmidt-Kaler, P. Walther and H. Weinfurter
for useful discussions and comments in both theoretical and experimental aspects of our work.
We acknowledge financial support from the European Commission 
(FETPRO QUIC (H2020-FETPROACT-2014
No. 641122), STREP EQuaM and STREP RAQUEL), 
the European Research Council (AdG OSYRIS and AdG IRQUAT), 
the Spanish MINECO (grants no.  FIS2008-01236, FIS2013-46768-P FOQUS, FISICATEAMO FIS2016-79508-P
and Severo Ochoa Excellence Grant SEV-2015-0522) with the support of FEDER funds, 
the Generalitat de Catalunya (grants no.~SGR 874, 875 and 966), CERCA  Program/Generalitat  de  Catalunya
and Fundaci{\'o} Privada Cellex.
AR also thanks support from the CELLEX-ICFO-MPQ fellowship.

\vspace{.7cm}
\noindent \textbf{Author contributions}
\vspace{.3cm}

M.~N.~B.\ and A.~R.\ have equally contributed to this work. 
M.~L.\ and A.~W.\ have supervised the project.
All authors discussed the results and contributed to the final manuscript.\\

\clearpage

\section*{Supplementary Information}

\section*{Supplementary Note 1: \\ Definitions of heat}\label{SupSubs:Heat}

In the main text, the heat dissipated in a process involving a system and a bath B 
has been defined as $\Delta Q = k T \Delta \mathcal{S}_\B$, such as 
the common description: ``\emphasize{flow of energy to a bath some way other than through work}'' suggests.
Note, however, that this is not the most extended definition of heat that one finds in many works, e.g., \cite{Reeb14_supp, Jennings10_supp}, where heat is defined as the change in the internal energy of the bath, i.~e.
\begin{equation}
  \Delta \tilde Q \coloneqq -\Delta E_\B\, ,
\end{equation}
and no different types of energy are distinguished in this increase of energy.
In this section, we compare these two definitions and argue why the approach taken here, though less extended, seems the most appropriate.

The ambiguity in defining heat comes from the different ways in which the change in the internal energy of the system $E_\Sys$ can be decomposed. More explicitly, let us consider a unitary process $U_\SB$ acting on
a system-bath state $\rho_\SB$ with $\rho_\B=\tr_\Sys \rho_\SB = \tau_\B \propto e^{- H_\B/kT}$ and 
global Hamiltonian $H=H_\Sys\otimes \id +\id \otimes H_\B$. 
The change in the total internal energy $\Delta E_\SB$ is the sum of system and bath internal energies
$\Delta E_\SB=\Delta E_\Sys +\Delta E_\B$, or equivalently
\begin{equation}
\Delta E_\Sys = \Delta E_\SB-\Delta E_\B\, .
\end{equation}
Many text-books identify in this decomposition $\Delta W\coloneqq-\Delta E_\SB$ as work and 
$\Delta \tilde Q= -\Delta E_\B$ as heat.
Nevertheless, note that it also assigns to heat increases of the internal energy that are not irreversibly lost and can be recovered when having a bath at our disposal.

To highlight the incompleteness of the above definition, let us consider a reversible process $U_\SB=\id \otimes U_\B$ that acts trivially on the system. 
Then, even though the state of the system is untouched in such a process, 
the amount of heat dissipated is 
$\Delta \tilde Q=-\Delta E_\B= \tr [H_\B(\rho_\B-U_{B}\rho_\B U_\B^\dagger)]$.

In order to avoid this kind of paradoxes and in the spirit of the definition given above,
we subtract from $\Delta E_\B$ its component of energy that can still be extracted 
(accessed). 
Then for a transformation $\rho_\B \rightarrow \rho_\B^\prime$, the heat transferred is given as
\begin{equation}\label{eq:diff-heat}
\begin{split}
 \Delta Q &= -(\Delta E_\B-\Delta F_\B), \\
%  \mathcal{S}(\rho_{B}^\prime \|\rho_{B} ) T, \\
          &= -kT\, \Delta \mathcal{S}_\B,
 \end{split}
\end{equation}
where $\Delta F_\B=F(\rho_\B^\prime)-F(\rho_\B)$ is the work stored on the bath 
and can be extracted. 
Here, $F(\rho_X)=E_X- kT\, \mathcal{S}(\rho_X)$ is the Helmholtz free energy, 
$E_X$ is the internal energy and 
% $S(\rho_{B}^\prime \|\rho_{B})= \mbox{Tr}\left[\rho_{B}^\prime (\log_2\rho_{B}^\prime - \log_2\rho_{B} ) \right]$ is the relative entropy and 
$\Delta \mathcal{S}_\B=\mathcal{S}(\rho_\B^\prime)-\mathcal{S}(\rho_\B)$ 
is the change in the bath's von Neumann entropy, 
$\mathcal{S}(\rho_\B) = -\mbox{Tr}\left[\rho_\B \log_2 \rho_\B \right]$. 
Throughout this work, we consider $\log_2$ as the unit of entropy.

Let us remark that in practical situations, in the limit of large baths, both definitions coincide. To see it, take Supplementary Eq.~\eqref{eq:diff-heat} and note that both definitions only differ in the free energy difference term, which together with the fact that the free energy is minimized by the thermal state, implies that the difference is very small when the bath is slightly perturbed. 
However, when studying thermodynamics at the quantum regime with small machines approaching the nanoscale such conceptual differences are crucial to extend, for instance, the domain of standard thermodynamics to situations where the correlations become relevant.

Note finally that both definitions express a \emphasize{path dependent quantity} of the system like heat in terms of a difference of state functions of the bath. 
The path dependence character comes from the fact that there are several processes that leave the system in
the same state but the bath in a different one. 
This connects with Clausius inequality, which is usually stated as
\begin{equation}
\oint \frac{\mathrm{d} Q}{T}\leqslant 0
\end{equation}
where the integral is taken over a cyclic path and the equality is only saturated
by quasiestatic processes.
In our framework and for the case of defining heat by means of information (entropy), 
the Clausius inequality is a consequence of the positivity of the mutual information.
That is, by assuming global entropy preservation we have
\begin{equation}
\Delta \mathcal{S}_\Sys=-\Delta \mathcal{S}_\B +\Delta I(\Sys:\B)=\frac{\Delta Q}{T} +\Delta I(\Sys:\B)
\end{equation}
where $I(\Sys:\B)=\mathcal{S}_\Sys+\mathcal{S}_\B-\mathcal{S}_\SB$ is the mutual information.
For an initially uncorrelated system-bath, the mutual information can only increase
$\Delta I(\Sys:\B)\ge 0$, and
\begin{equation}
\frac{\Delta Q}{T} \leqslant \Delta \mathcal{S}_\Sys\, .
\end{equation}

For the definition of heat as an increase of the internal energy, we have
\begin{equation}
\Delta \tilde Q= \Delta Q -\Delta F_\B\leqslant \Delta Q\, ,
\end{equation}
where we have used Supplementary Eq.~\eqref{eq:diff-heat} and the positivity of the free energy change.
In sum, for the case of initially uncorrelated states, we 
recover the \emphasize{Clausius inequality},
\begin{equation}\label{eq:Clausius}
\Delta \tilde Q \leqslant \Delta Q \leqslant T \Delta S\, .
\end{equation}
The deficit for the first inequality to be saturated is $\Delta F_\B$, that is, the energy that can still extracted from the bath. 
If one has a limited access to the bath,
an apparent relaxation process will follow and the bath will thermalize keeping its energy constant.
This will imply an entropy increase of the bath $\Delta F_\B/T$ which will make
$\Delta \tilde Q$ and $\Delta Q$ coincide.

The deficit to saturate the second inequality in Supplementary Eq.~\eqref{eq:Clausius} is $\Delta I(\Sys:\B)$, that is, the amount of enabled correlations during the process. One of the main ideas of this work is to show that these correlations capture a free energy that can be extracted.

\section*{Supplementary Note 2: \\ Set of operations}
The set of operations that we consider in this manuscript is the so called \emphasize{entropy preserving operations}.
Given a system initially in a state $\rho$, the set of entropy preserving operations are all the operations
that change arbitrarily the state but keep its entropy constant
\begin{equation}
\rho \to \sigma \hspace{.3cm}: \hspace{.3cm} \mathcal{S}(\rho)=\mathcal{S}(\sigma)\, ,
\end{equation}
where $\mathcal{S}(\rho)\coloneqq - \tr(\rho \log \rho)$ is the Von Neumann entropy.
It is important to note that an operation that acting on $\rho$ produces a state with the same entropy
does not mean that will also preserve entropy when acting on other states.
In other words, such entropy preserving operations are in general not linear, since
they have to be constraint to some input state.
In fact, in \cite{Hulpke2006}, it is shown that a quantum channel $\Lambda(\cdot)$ that
preserves entropy and respects
linearity, i.~e.\ $\Lambda(p \rho_1+ (1-p)\rho_2)=p\Lambda(\rho_1)+ (1-p)\Lambda(\rho_2)$,
has to be necessarily unitary.

One could think then that the extension of the unitaries to a set of entropy preserving operations 
is rather unphysical since they are not linear. 
However, they can be microscopically described by global unitaries
in the limit of many copies \cite{Sparaciari16}.
That is, given any two states $\rho$ and $\sigma$ with equal entropies
$S(\rho)=S(\sigma)$, then there exists a unitary $U$ and an additional system 
of $O(\sqrt{n\log n})$ ancillary qubits such that
\begin{equation}\label{eq:entropy-pres-micro}
\lim_{n\to \infty}
\|\tr_{\textrm{anc}}\left(U \rho^{\otimes n}\otimes\eta U^{\dagger}\right)-\sigma^{\otimes n}\|=0  \, ,
\end{equation}
where $\|\cdot\|$ is the one-norm and the partial trace is performed on the ancillary qubits. 
The reverse statement is also true, i.~e.\ if two states can be related as in Supplementary Eq.~\eqref{eq:entropy-pres-micro} then they have equal entropies.
This is proven in Theorem 4 of Ref.~\cite{Sparaciari16}.

Sometimes it can be interesting to restrict entropy preserving operations to also be  energy preserving. The set of energy and entropy preserving channels can also be described as a global energy preserving unitary in the many copy limit.
More explicitly, in Theorem 1 of Ref.~\cite{Sparaciari16}, it is proven that two states $\rho$ and $\sigma$ having equal entropies and energies ($S(\rho)=S(\sigma)$ and $E(\rho)=E(\sigma)$) is 
equivalent to 
the existence of some $U$ and an additional system A with $O(\sqrt{n\log n})$ ancillary qubits with Hamiltonian $\norm{H_\A}\leqslant O(n^{2/3})$ in some state $\eta$ for which Supplementary Eq.~\eqref{eq:entropy-pres-micro} is fulfilled.
Note that the amount of energy and entropy of the ancillary system per copy vanishes  in the large $n$ limit.

In sum, considering the set of entropy preserving operations 
means implicitly taking the limit of many copies and global unitaries.
In addition, as that the set of entropy preserving operations contains the set of unitaries, 
any constraint that appears as a consequence of entropy preservation will be 
also respected by individual quantum systems.

The Hamiltonians of the system and the bath are the same before 
and after the transformation $\Lambda(\cdot)$.
This can be done without loss of generality since, when this is not the case
and the final Hamiltonian is different from the initial one,
the two situations are related by a simple quench (instantaneous change of the Hamiltonian).
More explicitly, let us consider a process (a) with equal initial and final Hamiltonian, and an identical process (b) with different ,
\begin{align}
\textrm{(a)} \hspace{1.5cm}&(H,\rho_{\textrm{i}}) \ \to \ (H,\rho_{\textrm{f}})\\
\textrm{(b)} \hspace{1.5cm}&(H,\rho_{\textrm{i}}) \ \to \ (H',\rho_{\textrm{f}})
\end{align}
where $\rho_{\textrm{i}/\textrm{f}}$ is the inital/final state, $H$ the inital Hamiltonian
and $H'$ the final Hamiltonian of the process with different Hamiltonians.
Then, it is trivial to relate the work and heat
involved in both processes
\begin{align}
W'&=W+\tr\left((H-H')\rho_{\textrm{f}}\right)\\
Q'&=Q \, ,
\end{align}
where $W'$ and $Q'$ are the work and heat associated to the process (b) and we have only used that the process (b) is the composition of the process (a) followed by a quantum quench.

Let us finally point out that initially and finally the Hamiltonians
of system and bath are not interacting, or in other words, the system is decoupled from the bath
\begin{equation}
H=H_\Sys\otimes \mathbb{I}+\mathbb{I}\otimes H_\B\, ,
\end{equation}
with $H_{S/B}$ the Hamiltonian of the system/bath.
This is a necessary condition to be able to consider system and bath as independent systems
each with a well defined notion of energy. Otherwise, assigning an energy to the system and to the bath would not be possible beyond the weak coupling limit. Note that the system and the bath interact (arbitrarily strongly) during the process, in which for instance a non-product unitary could be performed.

\section*{Supplementary Note 3: \\ The Landauer principle}
%: \protect\\ connecting heat and information}
\label{SupSubs:LEP}

The information theory and statistical mechanics have longstanding and intricate relation. In particular, to exorcise Maxwell's demon in the context of statistical thermodynamics, Landauer first indicated that information is physical and any manipulation of that has thermodynamic cost. As put forward by Bennett \cite{Bennett82_supp}, the Landauer information erasure principle (LEP) implies that 
``any logically irreversible manipulation of information, such as the erasure of a bit or the merging of two computation paths, must be accompanied by a corresponding entropy increase in non-information-bearing degrees of freedom of the information-processing apparatus or its environment.''

Following the definition of heat, it indicates that, in such processes, \emphasize{ entropy increase in non-information-bearing degrees of freedom} of a bath is essentially associated with flow of heat to the bath. The major contribution of this work is to exclusively quantify heat in terms of flow of information, instead of counting it with the flow of non-extractable energy, the work. To establish this we start with the case of information erasure of a memory. Consider a physical process where an event, denoted with $i$, happens with the probability $p_i$. Then storing (classical) information memorizing the process means constructing a $d$-dimensional system (a memory-dit) in a state $\rho_\Sys=\sum_i p_i |i \rangle \langle i |$, where $\{ |i \rangle \}$ are the orthonormal basis correspond to the event $i$. In other words, memorizing the physical process is nothing but constructing a memory state $\rho_\Sys=\sum_i p_i |i \rangle \langle i |$ from a memoryless state $|i \rangle \langle i |$ where $i$ could assume any values $1 \leqslant i \leqslant d$. On the contrary, process of erasing requires the transformation of a memory state $\rho_\Sys=\sum_i p_i |i \rangle \langle i |$ to a memoryless state $|i \rangle \langle i |$ for any $i$. Landauer's erasure principle (LEP) implies that erasing information, a process involving a global evolution of the memory-dit system and its environment, is inevitably associated with an increase in entropy in the environment. 

In establishing the connection between information erasing and heat dissipation, we make two assumptions to start with. First, the memory-system (S) and bath (B) are both described by the Hilbert space $\mathcal{H}_\Sys \otimes \mathcal{H}_\B$. Secondly, the erasing process involves entropy preserving operation $\Lambda^{\SB}$, i.e.,  $\rho_\SB^\prime= \Lambda^{\SB} \left( \rho_\SB \right)$. The latter assumption is most natural and important, as it preserves information content in the joint memory-environment system. Without loss of generality, one can further assume that the system and bath Hamiltonians remain unchanged throughout the erasing process, to ease the derivations.  

Now we consider the simplest information erasing scenario, which leads to LEP in its 
traditional form. In this scenario, a system $\rho_\Sys$ is brought in contact with a 
bath $\rho_\B$ and the system is transformed to a information erased state, say $|0\rangle \langle 0|_\Sys$, by performing a global entropy-preserving operation $\Lambda^{\SB}$, i.e.,
\begin{align}
 \rho_\Sys \otimes \rho_\B \xrightarrow{\Lambda^{\SB}} |0\rangle \langle 0|_\Sys \otimes \rho_\B^\prime,
\end{align}
where initial and final joint system-bath states are uncorrelated. The joint operation guarantees that the decrease in system's entropy is exactly equal to the increase in bath entropy and heat dissipated to the bath is $\Delta Q=-kT\, \Delta \mathcal{S}_\B$. It clearly indicates that an erasure process is expected to \emphasize{ heat up} the bath. 
This in turn also says that $\Delta Q=kT\, \Delta \mathcal{S}_\Sys$, where $\Delta \mathcal{S}_\Sys=\mathcal{S}(\rho_\Sys^\prime)-\mathcal{S}(\rho_\Sys)$. In the case where the $d$-dimensional system memorizes maximum information, or in other words it is maximally mixed and contains $\log_2 d$ bits of information, 
the process dissipates an amount $kT \log_2 d $ of heat to completely erase the information. 
In other words, to erase one bit of information system requires the dissipation of $kT$ 
of heat and we denote it as one \emphasize{ heat-bit} or \emphasize{ $\ell$-bit} (in honour of Landauer).

In the case where the final state may be correlated, the dissipated heat in general
is lower bounded by the entropy reduction in the system, i.e.,  
\begin{align}
 |\Delta Q| \geqslant  kT\, |\Delta \mathcal{S}_\Sys|.
\end{align}
This is what is generally known as the Landauer's erasure principle (LEP), in terms of heat. 

The above formulation of LEP crucially relies on the fact that any change 
in system entropy leads to a larger change in the bath entropy, which is also 
traditionally known as the \emphasize{ second law} for the change in the information, i.e.,
\begin{align}
 \Delta \mathcal{S}_\B \geqslant -\Delta \mathcal{S}_\Sys.
\end{align}
% $\Delta \mathcal{S}_\B \geqslant -\Delta \mathcal{S}_\Sys$. 
However, it is limited by the assumptions made above and can be violated with initial correlations. 
Consider the examples in section \ref{SupSec:violation} of the Supplementary Information. 
In both the examples, $\Delta \mathcal{S}_\B \ngeq -\Delta \mathcal{S}_\Sys$. Therefore, one has to replace it with universal informational second law.

\section*{Supplementary Note 4: \\ Violations of laws of thermodynamics}\label{SupSec:violation}
In order to highlight how the laws of thermodynamics break down in the
presence of correlations, let us discuss the following two examples. 
In the first, the system S is purely classically correlated with the 
bath B at temperature $T$, while in the other they are jointly in a 
pure state and share quantum entanglement. 
In both the examples the Hamiltonians of the system and bath ($H_\Sys$ and $H_\B$) 
remain unchanged throughout the processes.

\noindent\medskip
\emphasize{\textbf{Example 1} -- Classical correlations.} 
\begin{align}
 \rho_\SB=\sum_i p_i \ket{i}\bra{i}_\Sys \otimes \ket{i}\bra{i}_\B\xrightarrow{U_\SB^c} \rho_\SB^\prime=\ket{\phi}\bra{\phi}_\Sys \otimes \sum_i p_i \ket{i}\bra{i}_\B, \nonumber
\end{align}

\noindent
\emphasize{\textbf{Example 2} -- Entanglement.} 
\begin{align}
\ket{\Psi}_\SB=\sum_i \sqrt{p_i} \ \ket{i}_\Sys \ket{i}_\B \xrightarrow{U_\SB^e} \ket{\Psi}_\SB^\prime = \ket{\phi}_\Sys \otimes \ket{\phi}_\B, \nonumber
\end{align}
where in both examples $\ket{\phi}_X=\sum_i \sqrt{p_i} \ \ket{i}_X$ with $X\in\{ \Sys,\B \}$
and $1>p_i\ge 0$ for all $i$.
Note that the unitaries, $U_\SB^c$ and $U_\SB^e$, leave the 
local energies of system and bath unchanged, and $U_\SB^c$ does not change the bath state.

\subsection{Violations of first law}

In \emphasize{ Example 1}, the Helmholtz free energy of the system increases 
$F(\ket{\phi}_\Sys)>F(\rho_\Sys)$ and therefore a work $-\Delta W_\Sys=\Delta F_\Sys>0$ is 
performed on the system. To assure the energy conservation of the system, an 
equal amount of heat is required to be transferred to the bath.
Surprisingly, however, no heat is transferred to the bath as it remains unchanged. 
Thus $\Delta E_\Sys \neq  -\Delta W_\Sys + \Delta Q$, i.e.~the energy conservation is 
violated and so the first law. 

A further violation can also be seen in \emphasize{ Example 2} involving system-bath 
quantum entanglement. In this case, a non-zero work $-\Delta W_\Sys=\Delta F_\Sys > 0$
has been performed on the system, and a heat flow to the bath is expected. 
In contrast, there is a \emphasize{ negative} heat flow to the bath! Therefore, it 
violates the first law, i.e.~$\Delta E_\Sys \neq -\Delta W_\Sys + \Delta Q$.

\subsection{Violations of second law and anomalous heat flows}

We now show how correlation could result in a violation of the \emphasize{ Kelvin-Planck} 
statement of the second law, which states: 
\emphasize{ No process is possible whose {\textbf \emphasize{ sole}} result is 
the absorption of heat from a reservoir and the conversion of this heat into work.}
In \emphasize{ Example 1}, no change in the local bath state indicates that there is 
no transfer of heat. However, the change in the Helmholtz free energy of the 
local system is $-\Delta W_\Sys=\Delta F_\Sys >0$. Thus, a non-zero amount of work is 
performed on the system without even absorbing heat from the bath ($\Delta Q=0$). 

The situation becomes more striking in \emphasize{ Example 2}, with initial system-bath 
entanglement. In this case,  $-\Delta W_\Sys=\Delta F_\Sys >0$ amount of work is performed 
on the system. However, not only is there no heat flow from the bath to the system, 
but there is a \emphasize{ negative} heat flow to the bath! Thus, the second law is violated.

We next see how the presence of correlations can lead to anomalous heat flows and 
thereby a violation of the \emphasize{ Clausius} statement based second law. 
Such violations were known for the other definition of heat $\Delta Q = - \Delta E_\B$ 
(see \cite{Jennings10_supp} and references therein). Here
we show that such violations are also there with new heat definition $\Delta Q = - kT\Delta S_\B$.
Let $\rho_\AB \in \mathcal{H}_\A \otimes \mathcal{H}_\B$ be an initial bipartite finite dimensional state whose marginals $\rho_\A=\tr_\B \rho_\AB= \frac{1}{Z_\A}\exp[-\frac{H_\A}{kT_\A}]$ and $\rho_\B= \frac{1}{Z_\B}\exp[-\frac{H_\B}{kT_\B}]$ are thermal states at different temperatures $T_\A$ and $T_\B$ and with Hamiltonians $H_\A$ and $H_\B$.
In absence of initial correlations between the baths A and B,
any energy preserving unitary will respect Clausius' statement of the second law. However, if
initial correlations are present, this will not be necessarily the case.

Consider a state transformation
$\rho_\SB^\prime=U_\AB \rho_\AB U_\AB^\dag$ where $U_\AB$ is a energy preserving unitary acting on $\rho_\AB$.
As the thermal state minimizes the free energy,  the final reduced states $\rho_\Sys^\prime$ and $\rho_\B^\prime$  have increased their free energy,
\begin{align}
\Delta E_\A -k T_\A \Delta S_\A &\ge 0 \label{eq:free-energy-increaseA}\\
\Delta E_\B -k T_\B \Delta S_\B &\ge 0\, ,\label{eq:free-energy-increaseB}
\end{align}
where $T_{\A/\B}$ is the initial temperature of the baths, and $\Delta E_{\A/\B}$ and $\Delta S_{\A/\B}$ are the change in internal energy and entropy respectively.

By adding Supplementary Eqs.~\eqref{eq:free-energy-increaseA} and \eqref{eq:free-energy-increaseB}, and considering energy conservation, we get
\begin{equation}\label{eq:Clausius-step}
T_\A \Delta S_\A + T_\B \Delta S_\B \le 0\, .
\end{equation}
Due to the conservation of total entropy,
the change in mutual information is simply $\Delta I(\A:\B)=\Delta S_\A+ \Delta S_\B$,
with $I(\A:\B)=\mathcal{S}_\A + \mathcal{S}_\B - \mathcal{S}_\AB$. 
This allows us to rewrite Supplementary Eq.~\eqref{eq:Clausius-step} in terms of only the entropy change in A as
\begin{equation}\label{eq:Clausius-real}
(T_\A - T_\B) \Delta S_\A  \le - T_\B \Delta I(\A:\B)\, .
\end{equation}
If the initial state $\rho_\AB=\rho_\A\otimes \rho_\B$ is uncorrelated, then the change in mutual information is necessarily positive $\Delta I(\A:\B)\ge 0$, and
\begin{equation}
k(T_\A - T_\B)\Delta S_\A =-\Delta Q_\A \frac{T_\A - T_\B}{T_\A} \le 0. \label{eq:clausius-formula}
\end{equation}
To see that this equation is precisely the Clausius statement, consider without loss of generality that A is the hot bath and $T_\A-T_\B>0$. Then, Supplementary inequality \eqref{eq:clausius-formula} implies an entropy reduction of the hot bath $\Delta S_\A\le 0$ i.~e.\ a heat flow from the hot bath to the cold one.

However, if the the system is initially correlated, the process can reduce 
the mutual information, $\Delta I(\A:\B) < 0$,
and Supplementary Eq.~\eqref{eq:Clausius-real} allows a heat flow from the cold bath to the hot one.

\subsection{Violations of zeroth law}

The zeroth law establishes the notion of thermal equilibrium as an equivalence relation, in which
temperature labels the different equivalent classes.
To see that the presence of correlations also invalidates the zeroth law, 
we show that the transitive property of the equivalence relation is
not fulfilled. 
Consider a bipartite system AC in an initial correlated state $\rho_{\AC}$, 
like in \emphasize{ Examples 1} and \emphasize{ 2},
and a third party B which is in a thermal state at the same temperature 
of the marginals $\rho_\A$ and $\rho_\textrm{C}$.
Then, while the subsystems AB and BC are mutually in equilibrium, the subsystems AC
are not, clearly violating transitivity.
There are several ways to realize that the parties AC are not in equilibrium.
One way is to see that any energy preserving unitary, except for the identity, decreases the amount of
correlations between the parties, $\Delta I(\A:\textrm{C})<0$, 
which implies that the initial state is not stable.
This can be shown from Supplementary Eq.~\eqref{eq:Clausius-step} for the particular case
of equal temperatures and the definition of mutual information.
Another way is to see that the Helmholtz free energy follows $F(\rho_{\AC}) > F(\rho_\A\otimes\rho_\textrm{C})$.

\subsection{Violations of Landauer's erasure principle}

Another thermodynamic principle that breaks down when correlations are present
is Landauer's erasure principle. Landauer postulated that in order to erase one 
bit of information in the presence of a bath at temperature $T$, an amount of heat 
needed to be dissipated is $kT \log 2$.
As noted in \cite{Reeb14_supp}, when the system is classically correlated, there exists 
erasing process which does not increase entropy of the bath (see \emphasize{ Example 1}). 
The situation becomes more striking when the system shares quantum entanglement 
with the bath. This is the case of \emphasize{ Example 2} with initial entanglement, 
where instead of increasing, an erasing process reduces the entropy of the bath 
and the bath is cooled down.

\end{document}